\documentclass[a4paper, 10pt]{article}

\usepackage{times,xcolor,amsmath,amssymb,epsfig,graphics,graphicx}
\begin{document}
\begin{center}
{\Large \bf Response to: [Comment on "Quantization of the damped harmonic oscillator" [Serhan et al, J. Math. Phys. 59, 082105 (2018)]"]}
\end{center}

\vspace{0.21cm}

\begin{center}
{\large M. Serhan, M. Abusini, Ahmed Al-Jamel, H.El-Nasser and Eqab M. Rabei}\\
Physics Department, Faculty of Science, Al al-Bayt University, P.O. Box 130040, Mafraq 25113, Jordan
\end{center}

\begin{abstract}
This is a response to a recently reported comment \cite{Fernandez19} on paper [J. Math. Phys.59, 082105 (2018)] regarding the quantization of damped harmonic oscillator using a non-Hermitian Hamiltonian with real energy eigenvalues. We assert here that the calculation of Eq.~(29) of \cite{Serhan18} is incorrect, and thus the subsequent steps via the Nikiforov-Uvarov method are affected, and the energy eigenvalues should have been complex. However, we show here that the Hermiticity of the Hamiltonian should be firstly achieved to make the correct transition from classical Hamiltonian to quantum counterpart, and this can be reached using the symmetrization rule. Applying the canonical quantization on the resulted Hermitian Hamiltonian and then using the Nikiforov-Uvarov method correctly, the energy eigenvalues will be real and exactly as given by Eq.~(35) of \cite{Serhan18}.
\newline
\end{abstract}
In \cite{Serhan18}, we arrived at the classical Hamiltonian:
\begin{equation}
\label{1}
H(y,p_{y})=\frac{p_{y}^{2}}{2m}+\frac{1}{2}m\omega ^{2}y^{2}+\frac{1}{2}.
\lambda yp_{y}.
\end{equation}
In its present shape, this Hamiltonian is not Hermitian and the energy eigenvalues are not guaranteed to be real as pointed out in \cite{Fernandez19}. We also reported in \cite{Serhan18} that the canonical quantization of this Hamiltonian, with the form as in Eq.~(\ref{1}) is given via solving  
the Schrodinger equation for this system is
\begin{equation}
\label{eq12}
 \frac{d^2\psi}{dy^2}-\frac{m\lambda y}{i\hbar}\frac{d\psi}{dy}+\frac{2m}{\hbar^2}\left(E-\frac{1}{2}m\omega^2 y^2\right)\psi(y)=0
\end{equation}
which we reported that it yields real eigenvalues given as \cite{Serhan18}
\begin{equation}
\label{E}
 E_n=\hbar \sqrt{(\omega^2-\frac{\lambda^2}{4})} (n+\frac{1}{2}),   n=0,1,2,3,...
\end{equation}
We assert here in this response that we made a miscalculations in the derivation of Eq.~(\ref{E}) based on Eq.~(\ref{eq12}) using the Nikiforov-Uvarov method. 
However, the true result with real energy eigenvalues can be achieved if re-consider the transition from classical Hamiltonian to its quantum counterpart before the canonical quantization takes place. If we visit Eq.~(\ref{1}) again, we see that the third term $yp_{y}$ is classically equivalent to $p_{y}y$ as will as $\frac{\left(yp_{y}+p_{y}y\right)}{2}$. According to symmetrization rule \cite{Shewell59} and many advanced quantum mechanics textbooks \cite{Liboff03}, such operator ordering ambiguity can be resolved by firstly symmetrizing and expressing it as normal ordered as $\frac{\left(yp_{y}+p_{y}y\right)}{2}$ before quantization takes place. 
In adopting the symmetrization rule, the Hamiltonian in Eq.~(\ref{1}) can be prepared to be Hermitian before canonical quantization takes place as 
\begin{eqnarray}
\label{12}
H(y,p_{y})&=&\frac{p_{y}^{2}}{2m}+\frac{1}{2}m\omega ^{2}y^{2}+\lambda\frac{\left(yp_{y}+p_{y}y\right)}{4} \nonumber \\
&=&\frac{H+H^\dagger}{2}.
\end{eqnarray}
To find the corresponding energy eigenvalues, we next quantize the system with this Hermitian operator, which yields
\begin{equation}
\label{eq13}
 \frac{d^2\psi}{dy^2}-\frac{m\lambda y}{i\hbar}\frac{d\psi}{dy}+\frac{2m}{\hbar^2}\left(E-\frac{1}{2}m\omega^2 y^2-i\hbar\frac{\lambda}{4}\right)\psi(y)=0
\end{equation}
Therefore, the implementation of  of NU method gives $\stackrel{\sim }{\tau }=-\frac{m\lambda y}{i\hbar}$, $\sigma=1$, $~\stackrel{\sim}{\sigma }=\frac{2m}{\hbar^2}(E-\frac{1}{2}m\omega^2 y^2-i\hbar\frac{\lambda}{4})$. Based on these new choices, we then have 
\begin{equation}
\label{eq18}
 \pi =\frac{m\lambda y}{i\hbar}\pm\sqrt{\alpha^2 y^2-\beta+k}
\end{equation}
where $\beta\equiv\frac{2m}{\hbar^2}(E-i\hbar\frac{\lambda}{4})$ and $\alpha^2\equiv\frac{m^2\omega^2}{\hbar^2}-\frac{m^2\lambda^2}{4\hbar^2}$. 
The correct formula for $\lambda_n$ is then
\begin{equation}
\label{lambda}
 \lambda_n=\frac{2mE}{\hbar^2}-\alpha+\frac{m\lambda}{2i\hbar},
\end{equation}
which gives the extra imaginary term $\frac{m\lambda}{2i\hbar}$. Also,
\begin{equation}
\label{eq23}
 \lambda_n=2n\alpha.
\end{equation}
Comparing  the new results for $\lambda_n$ as given by Eq.~(\ref{lambda}) with Eq.~(\ref{eq23}), we obtain the energy eigenvalues as
\begin{equation}
\label{E2}
 E_n=\hbar \sqrt{(\omega^2-\frac{\lambda^2}{4})} (n+\frac{1}{2}),~~~n=0,1,2,3,...
\end{equation}
which are real.

To double check our results, we follow the procedure illustrated in \cite{Zafar02,Zafar19} to find the energy eigenvalues. By noting that
\begin{eqnarray}
\label{12b}
H(y,p_{y})&=&\frac{p_{y}^{2}}{2m}+\frac{1}{2}m\omega ^{2}y^{2}+\lambda\frac{\left(yp_{y}+p_{y}y\right)}{4} \nonumber \\
&=&\frac{\left(p_{y}+m\lambda y/2\right)^{2}}{2m}+\frac{1}{2}m\left(\omega ^{2}-\frac{\lambda^2}{4}\right) y^{2}
\end{eqnarray}
and that the corresponding eigenvalue equation is $H\psi=E\psi$, then $\eta H \eta^{-1} \eta \psi =E\eta \psi$. With the choice $\eta=\exp\left(i m\lambda y^2/4\hbar\right)$, we have 
\begin{eqnarray}
\label{15}
e^{im\lambda y^2/4\hbar}\left[p_{y}+m\lambda y/2\right]e^{-im\lambda y^2/4\hbar}&=&p_y \nonumber \\
e^{im\lambda y^2/4\hbar}\left[p_{y}+m\lambda y/2\right]^n e^{-im\lambda y^2/4\hbar}&=&(p_y)^n
\end{eqnarray}
Therefore,
\begin{equation}
\label{17}
\eta H \eta^{-1} \eta =\frac{p_{y}^{2}}{2m}+\frac{1}{2}m\left(\omega ^{2}-\lambda^2/ 2\right) y^{2}.
\end{equation}
The quantization of Eq.~(\ref{17}) is clearly the real energy eigenvalues as given by Eq.~(\ref{E}). 

\end{document}